\documentclass[a4paper,11pt]{article}

\usepackage{jinstpub} 

\usepackage{lineno}

\title{\boldmath Radiation studies performed on the High Luminosity ATLAS TileCal link Daughterboard}

\author[a,1]{E. Valdes Santurio\note{Corresponding author.}}
\author[a]{, S. Silverstein}
\author[a]{, C. Bohm}
\author[a]{, H. Motzkau}
\author[a]{, C. Clement}
\author[a]{, K. Dunne}
\author[a]{, S. Lee}
\author{on behalf of the ATLAS Tile Calorimeter System{$^{\textit{2,3}}$}\note{This work was supported by Stockholm University and CERN.} \note{\textbf{Copyright 2020, CERN, for the benefit of the ATLAS Collaboration. CC-BY-4.0 license.}}}

\affiliation[a]{Stockholm University, Stockholm, Sweden}

\emailAdd{eduardo.valdes@fysik.su.se}

\abstract{The new electronics of the ATLAS Tile Calorimeter for the HL-LHC interfaces the on-detector and off-detector electronics by means of a Daughterboard. The Daughterboard is positioned on-detector featuring commercial SFPs+, CERN GBTx ASICs, ProASIC FPGAs and Kintex Ultrascale FPGAs. The design minimizes single points of failure and mitigates radiation damage by means of a double redundant scheme, Triple Mode Redundancy, Xilinx Soft Error Mitigation IP, CRC/FEC for link data transfer, and SEL protection circuitry. We present an updated summary of the TID, NIEL and SEE qualification tests, and performance studies of the Daughterboard revision 6 design.}

\keywords{Radiation damage to electronic components, Front-end electronics for detector readout.}

\proceeding{Topical Workshop on Electronics for Particle Physics - TWEPP2022\\
  19 - 23 September, 2022\\
  Bergen, Norway}

\begin{document}

\maketitle

\flushbottom

\section{Introduction}
\label{sec:introduction}

The Tile Calorimeter (TileCal) is a sampling calorimeter of the ATLAS Experiment \cite{bib_atlas}. TileCal is in charge of identifying and measuring hadronic jets and transverse missing energy. The TileCal electronics will be upgraded to face the higher radiation levels and increased rates of pileup at the High Luminosity Large Hadron Collider (HL-LHC) \cite{bib_tilecal_phase_ii_tdr:2017}. The upgrade project aims to provide continuous digital read-out of all the TileCal with improved timing and better energy resolution.

The off-detector electronics of the upgraded TileCal will provide digitized signals at 40 MHz to the trigger systems through the Trigger and Data Acquisition interface (TDAQi), while the so-called Front-End Link eXchange (FELIX) system will read-out data stored in pipelines in Tile Preprocessors (TilePPr) at 1 MHz. The TilePPrs will interface the on- and off-detector electronics through multi-Gbps optical links. The power will be monitored and distributed to the on-detector electronics by the Detector Control System (DCS) that interfaces with the Low Voltage Power Supplies (LVPS) in charge of distributing power to all the digital and analogue electronics, and the High Voltage Power Supplies that will provide high voltage to each PMT.

\begin{figure*}[!hb]
	\centering
	\includegraphics[width=1.0\linewidth]{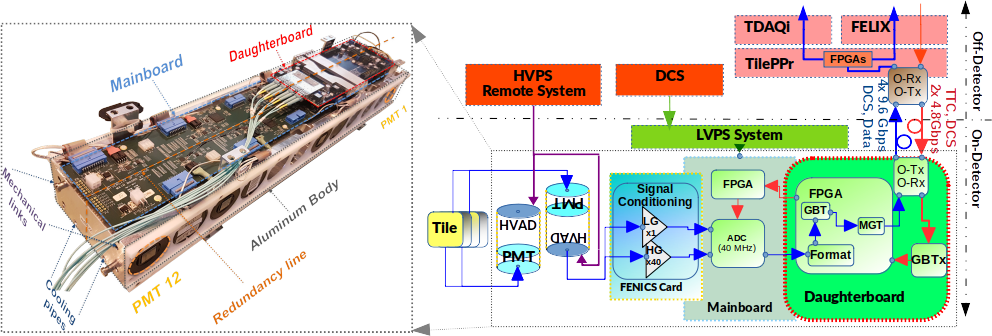}
	\caption{(Left:) TileCal Phase-II Upgrade Minidrawer. (Right:) On-Detector and Off-detector electronics block diagram.}
	\label{fig:tilecal_md_phase_ii}
\end{figure*}	

The on-detector is comprised by 896 Minidrawers (MDs), in which data from all TileCal PMTs will continuously be sampled at 40 MHz. Each MD (Figure: \ref{fig:tilecal_md_phase_ii}) will host up to twelve channels by means of twelve PMTs that turn light pulses to electric signals to be shaped and conditioned by Front-End Boards. A Mainboard (MB) will continuously sample and digitize two gains of shaped signals of the MD PMTs. A Daughterboard (DB) is in charge of distributing LHC synchronized timing, configuration and control to the front-end, and continuously transmitting the digital data from all the MB channels to the off-detector systems via multi-Gbps optical links.

\section{ATLAS TileCal link Daughterboard}
\label{sec:db6}
\begin{figure}[htbp]
\centering
\includegraphics[width=0.9\linewidth]{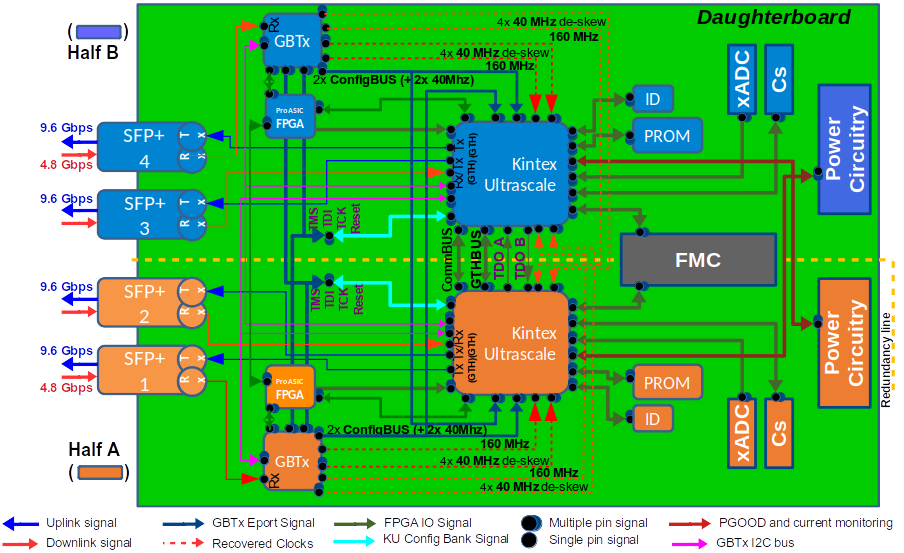}
\caption{The Daughterboard revision 6 block diagram.}
\label{fig:db6_block_diagram}
\end{figure}

The DB is a radiation tolerant board with a redundancy layer of two functionally-equivalent halves, interconnecting the on- and off-detector by means of a 400-pin FMC connector and multi-Gbps optical links, respectively \cite{bib_db}. The DB, currently on its sixth revision (DB6, Figure \ref{fig:db6_block_diagram}), powers the optical links by means of four SFP+ modules that drive two downlinks and four uplinks.

Taking into consideration the position of the DB in ATLAS, commercial off-the-shelf components have been used for the design following the availability of the required components in the market, with the exception of the CERN radiation hard GBTx ASICs \cite{bib_gbtx}, due to the importance of the stable clock recovery and distribution very specific to the LHC readout systems. Reported radiation tests studies performed in the ProASIC \cite{bib_proasic} and the Kintex Ultrascale FPGAs \cite{bib_ku_rad_1} \cite{bib_ku_rad_2} in addition to the tests performed internally in the collaboration were taken into consideration for the selection of the aforementioned components due to their complex nature. The remaining part of the COTS were tested up the the radiation requirements in parallel to the board development process. 


The information received by the board from the off-detector by means of 4.8 Gbps GBT-FEC downlinks powered by CERN radiation hard GBTx ASICs includes clock signals, time synchronization, and configuration commands (via the ConfigBUS path). Two ProASIC FPGAs buffer the resets and JTAG signals (TMS,TCK and TDI) received from the GBTx chips for the remote reconfiguration of the Xilinx Kintex Ultrascale (KU) FPGAs. The two 20 nm planar featured KU FPGAs distribute clocks and configuration received from the GBTx to the MB, while reading slow integrator data and fast digitized data from two different gains of the PMTs. The xADCs of the KU FPGAs sample temperature, voltage stability and currents. Additionally, the KU FPGAs are connected to the Cs and xADC headers for extra digital and analog capabilities respectively, and communicate with each other for monitoring purposes via the so called CommBUS and GTHBUS.
An over-current protection circuit triggers a power cycle in the case of the presence of high currents. Each  KU FPGA sends data off-detector by means of two copies of GBT-FEC protected words through two 9.6 Gbps uplinks. The board radiation tolerance is achieved by the redundancy layers in the design, using the Xilinx Soft Error Management (SEM), Triple Mode Redundancy (TMR), GBT-FEC protocol, and GBTx ASICs.

\section{Radiation requirements for the DB6}
\label{sec:rad-reqs}

The radiation levels that the DB is expected to be exposed to were calculated from the radiation simulations performed by the ATLAS Collaboration to estimate the expected radiation levels for the HL-LHC era \cite{bib_atlas_radiation}. The most irradiated positions where the DB will be installed was used as a reference value for the radiation requirements. Approximately $108$ Gy is the Total Ionizing Dose (TID) required for the design to be qualified, taking into consideration Safety Factors (SF) of $1.5$ for simulation and $3.0$ for lot variation in the components. The Non-Ionizing Energy Losses (NIEL) fluence required is for qualification is $13.16 \times 10^{12}$ n/cm$^{2}$, including SFs of $1.5$ for simulation, $1.3$ when using a non-monoenergetic neutron source, and $3.0$ to account for lot variation of the components. The expected fluence for Single Event Effects is $4.54 \times 10^{11}$ h/cm$^{2}$, where a SF = $3.0$ for simulation was included in the calculations. All the aforementioned requirement values have been put into place in accordance with the ATLAS Radiation Tolerant Electronics Proposed Guideline on COTS Lot to Lot Variation \cite{bib_retf}.

\section{Radiation tests on the DB6}
\label{sec:rad-tests}

Four Daughterboards were produced and separately used to test for radiation tolerance, labelled in this document as DB6-0, DB6-1, DB6-2 and DB6-3.

\begin{figure}[htbp]
\centering 
\includegraphics[width=1\textwidth]{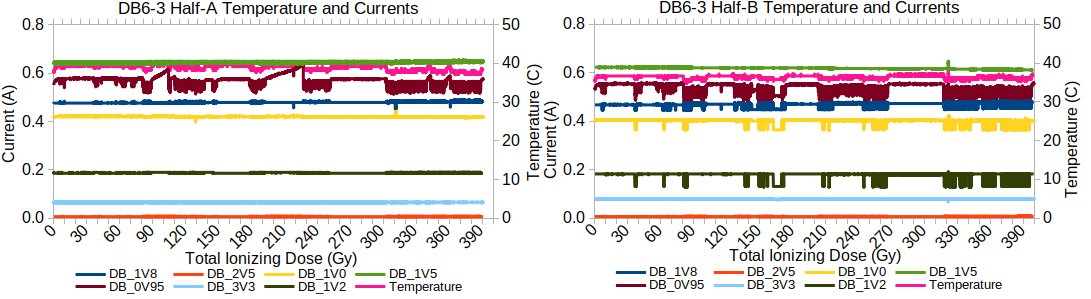}
\caption{\label{fig:db6_3_tid} TID test of DB6-3. All board currents and the temperatures of both KU FPGAs were monitored over the whole irradiation period delivered by a 54 MeV proton beam. }
\end{figure}

Two separate tests were performed to cover tolerance to TID. The KU FPGA temperatures and all the currents corresponding to all the point of load regulators of the irradiated boards were monitored. Special attention was placed on the total of ten Coretek SFP+, the six KU FPGAs, the six Microsemi ProaSIC3 FPGAs and the IS25LP128F reconfiguration FLASH memories. None of the components were damaged by the deposited TID:
DB6-1 and DB6-2 were irradiated with Gamma radiation at $^{60}$Co at the CERN CC60 facilities to 220 Gy at a rate of 3.37 Gy/h (Figure 9 of Ref.\cite{bib_db6_icaleps}) and up to 54 Gy at a rate of 0.33 Gy/h (Figure 10 of Ref. \cite{bib_db6_icaleps}), respectively. DB6-1 ran stable during the full test, however it showed an increase in the current measured in the 0.95 V power line of both sides of the board of at around 140 Gy. This was demonstrated to be correlated with the failure of the active components of the fan used to keep the KU FPGAs cool, leading to the increase of temperature on the FPGA. The currents, temperatures and functionalities of the DB6-2 were stable during the full run.
DB6-3 was irradiated in the AlC-144 cyclotron proton beam line with 54 MeV protons at the IFJ in Krakow to 381 Gy for half-A FPGA and 400 Gy for half-B FPGA at a rate of 2.88 Gy/h (Figure \ref{fig:db6_3_tid}). All the currents and the temperatures were stable during the full run. The higher fluctuations compared with DB6-1 (0 to 140 Gy) and DB6-2 are due to active Single Event Upset (SEU) corrections by the Xilinx SEM and resets applied when an uncorrectable SEU appeared during the run.

For the Single Event Latchups (SEL) tests, there was no prototype of the DB produced and the design was migrating from the 16 nm FinFET powered KU+ FPGA to the KU FPGA, due to the KU+ alternative been susceptible to SEL occurrences. As an alternative to the ProASIC, a Lattice ICE40LP FPGA was also tested \cite{bib_db6_icaleps}. Two separate tests were performed for SEL, one as described for the DB6-3 with 54 MeV protons, where the FPGAs were irradiated to a fluence of $2.62 \times 10^{11}$ p/cm$^{2}$. In the second test, two Trenz TE0841 boards (each with the same KU FPGA used in the DB design), a Lattice ICE40LP FPGA and a ProASIC3E A3P31500 FPGA were irradiated with 226 MeV protons to a fluence of $1.11 \times 10^{12}$ p/cm$^{2}$, for which all the currents of the FPGAs were monitored (Figure 6 of Ref. \cite{bib_db6_icaleps}). No SEL were observed in the units under test in either of the tests.

A set of tests was arranged in coordination with the RIC Jozef Stefan Institute in Ljubliana, to cover the NIEL tolerance, where a DB (DB6-0) and a set of test boards with non-radiation qualified DB components were irradiated in the beam port number 6 of the TRIGA reactor \cite{bib_triga_reactor} to a fluence of $14 \times 10^{12}$ (1 MeV equivalent n)/cm$^{2}$.
The post-radiation tests performed on DB6-0 demonstrated the radiation tolerance of the KU FPGAs up to the exposed fluence. In the case of the ProASIC, some after-effects could be seen where the firmware was not working as expected and re-configuration was not possible. However, the firmware functionality of all devices was recovered after annealing, with re-configuration capability recovered all of the devices. The same applies to the IS25LP128F FLASH memories, where after annealing some of the devices recovered the re-programing capabilities, with all the contents of the memory not being affected in any of the devices.

Four different variants of 10 Gbps multi-mode SFP+ devices were tested: AVAGO, CORETEK, FS-Europe and Direktronik. The two AVAGO AFBR-709SMZ devices completely failed to pass the irradiation tests done up to $9 \times 10^{12}$ (1 MeV equivalent n)/cm$^{2}$ fluence. One out of nine CORETEK CT-A000NPP-SB1L-D devices tested failed to function after irradiating to $9 \times 10^{12}$ (1 MeV equivalent n)/cm$^{2}$ fluence. In the case of the Direktronik 33-4248 and the FS-Europe SFP-10GSR-85 devices, one of each of the two brands was exposed to a higher fluence of $14 \times 10^{12}$ (1 MeV equivalent n)/cm$^{2}$, and a batch of ten of each of the two brands was exposed to $7 \times 10^{12}$ (1 MeV equivalent n)/cm$^{2}$ fluence. All of the samples irradiated from Direktronik and FS-Europe were fully functional after irradiation.

\begin{figure}[htbp]
\centering 
\includegraphics[width=1\textwidth]{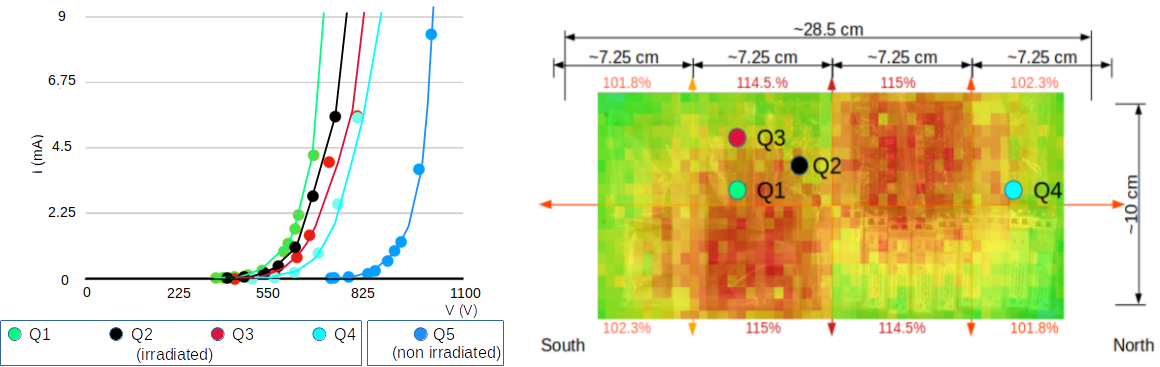}
\caption{\label{fig:db6_mosfet} (Left:) Threshold shift depicted in an IV curve of four of the irradiated FDFMA3N109 MOSFETs. (Right:) Position of the irradiated MOSFETs in the flux 2D slice map of the reactor cavity.}
\end{figure}

It was seen that the threshold voltages for the FDFMA3N109 MOSFETs used in the design shifted with the exposure (Figure \ref{fig:db6_mosfet}) due to the Gamma component present in the reactor cavity (195 Gy).
The asymmetrical profile of the flux in the cavity was used to have a perception of the correlation between the shifts of the thresholds in the MOSFETs and the exposed fluence.
Four out of fifteen QUAD oscillators LTC6902IMS irradiated failed to pass the post irradiation tests, therefore the device needs to be removed from the design.
The rest of the components (DAN217T146CT Schottky diodes, 540BAA100M000BBG oscillators and INA333AIDRG instrumentation amplifiers) passed the post radiation tests with no issues.

\section{Conclusions}
\label{sec:conclusions}

Radiation tests were conducted to determine the suitability of each component and part of the DB6 design for meeting HL-LHC requirements. The results of the completed tests have demonstrated that the key components of the design have TID, NIEL, and SEL resistance up to the necessary radiation levels, with a few non-critical, radiation-sensitive parts still requiring attention. Additional TID and SEL testing needs to be carried out on certain SFP+ variants.

Some radiation sensitive parts of the features of the design need modification:
\begin{itemize}
  \item The supplementary over-current protection circuit added for SEL mitigation has to be removed because of the radiation induced variations in the threshold of the FDFMA3N109 MOSFETS used in the design.
  \item The LTC6902IMS QUAD oscillators, previously used to de-couple the phase of the DC-DC converters, has to be removed from the design. Further research is necessary to evaluate the effect of this design modification on board noise levels.
\end{itemize}
The removal of the aforementioned MOSFETs and QUAD oscillatorswill make the design components at a board level sufficiently resistant to the expected TID, NIEL ans SEL irradiation levels.

The development roadmap for the DB R\&D project includes creating additional prototypes incorporating the previously mentioned modifications, conducting SEU tests with operational firmware to evaluate the impact of SEU rates on board performance, and analyzing test results to determine the final selection of the SFP+ variant that will be integrated to the final board design. Once the design is adjusted based on the test outcomes, prototypes will be manufactured and tested as part of the production phase. The project aims to produce a total of 930 DB6 units as Stockholm University's contribution to the HL-LHC era for TileCal.

\end{document}